\documentclass[11pt,twocolumn]{article}
\usepackage{amsmath,epsfig,psfig,url,graphicx,xspace}

\renewcommand{\footnotesize}{\small}

\makeatletter
{\catcode`\/\active\catcode`\_\active\catcode`\.\active
\gdef\URLslash{\futurelet\next\@@URLslash}%
\gdef\@@URLslash{\ifx\next\URLslash\char`\/\else\slash\fi}%
\gdef\URLdot{\char`\.\penalty\exhyphenpenalty}%
\gdef\URLprepare{\catcode`\/\active\catcode`\_\active\catcode`\.\active
        \let/\URLslash\let.\URLdot\def~{\char`\~}\def_{\char`\_}}}%
\def\URL{\bgroup\URLprepare\realURL}%
\def\realURL#1{\tt #1\egroup}%

\def\Section {\S}

\newsavebox{\sbrack}
\newsavebox{\mbrack}
\newsavebox{\lbrck}
\newsavebox{\mlbrck}

\title{\sffamily\fontsize{18}{18}
  \textbf{Tycoon: a Distributed Market-based \\ Resource Allocation System}}

\newcount\aucount
\newcount\originalaucount
\newdimen\auwidth
\newdimen\auskip
\newcount\auskipcount
\newdimen\auskip
\global\auskip=1pc
\newdimen\allauboxes
\allauboxes=\auwidth
\newtoks\addauthors
\newcount\addauflag
\global\addauflag=0 

\gdef\numberofauthors#1{\global\aucount=#1
\ifnum\aucount>4\global\originalaucount=\aucount \global\aucount=4\fi 
\global\auskipcount=\aucount\global\advance\auskipcount by 1
\global\multiply\auskipcount by 2
\global\multiply\auskip by \auskipcount
\global\advance\auwidth by -\auskip
\global\divide\auwidth by \aucount}

\newfont{\affname}{phvr at 10pt}
\newfont{\eaddrfnt}{phvr at 8pt}

\author{
Kevin~Lai~~~~~~Bernardo~A.~Huberman~~~~~~Leslie~Fine\\
{\{klai, bernardo.huberman, leslie.fine\}@hp.com}\\\\
HP Labs
}

\global\def\@maketitle{%
  \newpage
  \begin{center}%
  \let \footnote \thanks
\expandafter\ifx\csname acmdescription\endcsname\relax
  \null
\else
  {\setbox0\hbox{\vbox{%
\begin{flushright}%
  \begin{tabular}[t]{r@{}}%
    \acmdescription
    \\ \noalign{\vskip0.25in}%
  \end{tabular}%
\end{flushright}%
\null}}\ht0=0pt\dp0=0pt\box0}%
\fi
    \vskip -1.2em%
    {\LARGE \@title \par}%
    \vskip 2em%
    {\large
      \lineskip .5em%
      \begin{tabular}[t]{c}%
        \@author
      \end{tabular}\par}%
    \vskip 1em%
    {\large \@date}%
  \end{center}%
  \par
  \vskip 1em}
\newcommand{\beq}{\begin{equation}}
\newcommand{\eeq}{\end{equation}}
\newcommand{\benq}{\begin{eqnarray}}
\newcommand{\eenq}{\end{eqnarray}}

\def\ged{\hbox{${\vcenter{\vbox{
        \hrule height 0.4pt\hbox{\vrule width 0.4pt height 6pt
        \kern5pt\vrule width 0.4pt}\hrule height 0.4pt}}}$}}

\def\compactify{\itemsep=0pt \topsep=0pt \partopsep=0pt \parsep=0pt}
 \let\latexusecounter=\usecounter
 \newenvironment{CompactItemize}
   {\def\usecounter{\compactify\latexusecounter}
    \begin{itemize}}
   {\end{itemize}\let\usecounter=\latexusecounter}

\begin{document}

\maketitle

\begin{sloppypar}

%
%

\begin{abstract}

P2P clusters like the Grid and PlanetLab enable in principle the same
statistical multiplexing efficiency gains for computing as the
Internet provides for networking. The key unsolved problem is resource
allocation. Existing solutions are not economically efficient and
require high latency to acquire resources. We designed and implemented
\emph{Tycoon}, a market based distributed resource allocation system
based on an \emph{Auction Share} scheduling algorithm. Preliminary
results show that Tycoon achieves low latency and high fairness while
providing incentives for truth-telling on the part of strategic users.

\end{abstract}

\section{Introduction}
\label{sec:introduction}

A key advantage of peer-to-peer clusters like the Grid and PlanetLab is
their ability to pool together computational resources that can then
be shared among peers. This allows increased throughput because of
statistical multiplexing and the fact that users have a bursty
utilization pattern. Sharing of nodes dispersed in the network
structure allows lower delay because applications can store data close
to potential users. Sharing allows greater reliability because of
redundancy in hosts and network connections. Finally, sharing allows
all of this at lower cost than a comparable private system.

The key problem for sharing resources in distributed systems is
allocation. Allocation is an issue because demand grows to fill the
available capacity. The resource demands of data mining, scientific
computing, rendering, and Internet services have kept pace with
hardware improvements. Problems in resource allocation are: the
existence of \emph{strategic} users who act in their own interests, a
rapidly changing and unpredictable demand, and hundreds or thousands
of unreliable hosts which are physically and administratively
distributed. In this paper we compare resource allocation systems on
the basis of: a) their \emph{economic efficiency}, which is the
percentage of \emph{surplus} that a system generates, where the
surplus for a transaction is the value of a resource to the
highest-valuing recipient minus the cost to the lowest-cost provider;
b) utilization, which is the percentage of resources used; latency,
which is the time to complete a task; c) risk, which is the lower
bound on the resources a task obtains in a time interval; and d)
fairness; which is the correlation between the actual distribution of
utility to users to the desired distribution.

The common approach to this problem is to use a Proportional
Share\cite{tijdeman1980} scheduler, where users have no incentive to
honestly report the value of their tasks. As a result, unimportant
tasks get as much resources as critical jobs. Without further
mechanisms, this causes economic efficiency to decrease as load
increases (see Figure~\ref{fig:market_no_market}), eventually going to
zero. To mitigate this, users engage in ``horse trading'' where one
user agrees not to run a unimportant jobs when another user is running
a critical one in exchange for the return favor in the future. This
process imposes excess latency and work on users. Another approach is
to use combinatorial optimization algorithms \cite{pinedo2002} to
compute a schedule. However, optimal algorithms are NP-hard and
share the Proportional Share problem of not eliciting users' true
value for tasks.

We use an economic approach based on auctions. Economic approaches
explicitly assume that users are strategic and that supply and demand
vary significantly over time. We use a \emph{strategyproof} mechanism
to encourage users to reveal their true need for a resource, which
allows the system to maximize economic efficiency.

However, existing economic resource allocation systems vary in how
they abstract resources. The Spawn system \cite{waldspurger1992}
abstracts resources as reservations for specific hosts at fixed
times. This is similar to the way that airline seats are
allocated. Although reservations allow low risk and low latency, the
utilization is also low because some tasks do not use their entire
reservations. Service applications (e.g., web serving, database
serving, and overlay network routing) result in particularly low
utilization because they typically have bursty and unpredictable
loads. An alternative is to combine Proportional Share with a market
mechanism \cite{chun2000}. This offers higher utilization than
reservations, but also higher latency and higher risk.

Our contributions are the design and implementation of
\emph{Tycoon}, a distributed market-based resource allocation
architecture, and of \emph{Auction Share}, a local resource
scheduler. \emph{Tycoon} distinguishes itself from other systems
in that it separates the allocation mechanism (which provides
incentives) from the agent strategy (which interprets
preferences). This simplifies the system and allows specialization
of agent strategies for different applications while providing
incentives for applications to use resources efficiently and
resource providers to provide valuable resources. \emph{Tycoon's}
distributed markets allows the system to be fault-tolerant and to
allocate resources with low latency. \emph{Auction Share} is the
local scheduling component of \emph{Tycoon}. As we show, it
distinguishes itself from market-based reservations and
proportional share by being high utilization, low latency, low
risk, and fair.

In \Section~\ref{sec:relatedwork}, we review related work in
cluster resource allocation and scheduling. In
\Section~\ref{sec:tycoon}, we describe the \emph{Tycoon}
architecture. In \Section~\ref{sec:auctionshare}, we describe the
\emph{Auction Share} scheduler.  We conclude in
\Section~\ref{sec:conclusion}.

\section{Related Work}
\label{sec:relatedwork}

In this section, we describe related work in resource allocation.
There are two main groups: those that consider the problem of
strategic users (the economic approach) and those that do not (the
computer science approach).

Examples of systems that use the economic approach are Spawn
\cite{waldspurger1992}, the Millennium resource allocator
\cite{chun2000}, and work by Stoica, et al. \cite{stoica1995}. These
systems are \emph{strategyproof}. A strategyproof mechanism forces
truth-telling to be the dominant strategy for each entity regardless
of the behavior of other entities. This ensures that rational entities
will tell the truth about their preferences (e.g., how important
particular resources are to them) and allow the overall system to be
economically efficient. As a result, these systems mitigate the effect
of strategic users. These systems differ in how they abstract
resources. Spawn uses a reservation abstraction that results in low
latency, low risk and low utilization (as described in
\Section~\ref{sec:introduction}. The Millennium resource allocator
uses Proportional Share, described in more detail below. Stoica, et
al. use a centralized priority queue of tasks that is not suitable for
a P2P system like PlanetLab. 

Proportional Share (PS) \cite{tijdeman1980} is one of the non-economic
resource allocators. Each PS process $i$ has a weight $w_i$. The share
of a resource that process $i$ receives over some interval $t$ where
$n$ processes are running is
\begin{equation}
\label{eq:proportionalshare}
\frac{w_i}{{\displaystyle\sum_{j=0}^{n-1} w_j}}.
\end{equation}
PS maximizes utilization because it always provides resources to needy
processes. Most recent work \cite{waldspurger1994} \cite{stoica1996}
on PS has focused on computationally efficient and fair
implementations, where fair is defined as having a minimal difference
between the actual allocation and the ideal one. We show in
\Section~\ref{sec:auctionshareresults} that the Auction Share (AS)
scheduler is fair and computationally efficient. One problem with PS
is its high risk, where risk is defined as the lower bound on the
resources a process can obtain in a time interval. A process's share
goes to zero as the sum of weights increases. Stoica, et
al. \cite{stoica1997} show that a PS scheduler can fix the shares of
processes that need controlled risk while varying the shares of other
processes. This is the approach we use in the AS scheduler to control
risk.

Another issue is latency. Any one process must wait for all the others
run. This delay makes no difference to a batch application like a
renderer, but could significantly affect a service application like a
web server. As described in \Section~\ref{sec:auctionshareresults}, PS
scheduling can increase latency by a factor of 10 even for four
processes. Borrowed Virtual Time (BVT) \cite{duda1999} schedulers are
a form of PS scheduler that addresses this problem. Our Auction Share
(AS) scheduler is similar to BVT in that it considers scheduling
latency and uses admission control to reduce risk. However, AS is a
simpler abstraction than BVT. BVT processes must specify three
variables: a warp value, a warp time limit, and an unwarp time
requirement. In contrast, AS processes only need to specify one: an
expected number of processor-seconds needed during an interval (the
interval is the same for all the processes if none are part of a
distributed application, as BVT assumes). Also, if an application
incorrectly sets its BVT parameters, it can significantly affect the
performance of other applications. In
\Section~\ref{sec:auctionshareresults}, we show that if an AS process
sets its parameter incorrectly, it only affects its own performance. 

Lottery scheduling \cite{waldspurger1994} is a PS-based abstraction
that is similar to the economic approach in that processes are issued
tickets that represent their allocations. Sullivan and Seltzer
\cite{sullivan2000} extend this to allow processes to barter these
tickets. Although this work provides the software infrastructure for
an economic mechanism\footnote{By \emph{mechanism} we mean the system
that provides an incentive for users to reveal the truth (e.g., an
auction)}, it does not provide the mechanism itself.

Similarly, SHARP \cite{fu2003} provides the distributed infrastructure
to manage tickets, but not the mechanism or bartering agent
strategies. In addition, SHARP and work by Urgaonkar, et
al. \cite{urgaonkar2002} use an overbooking resource abstraction
instead of PS. An overbooking system promises probabilistic resources
to applications. Overbooking has essentially the same risk as PS
because in the worst case, unexpected demand can deprive an
application of its resources.

Another class of non-economic algorithms are those based on
combinatorial optimization \cite{papadimitriou1982}
\cite{pinedo2002}. This approach assumes that the load is
deterministic and uses a centralized NP-hard algorithm to compute the
optimal algorithm. As a result, it would perform poorly with the
rapidly changing and unpredictable loads typical on the Grid and
PlanetLab. In addition, the centralized optimizer would impose
bottlenecks and decrease the reliability of an otherwise decentralized
system.

\section{Tycoon}
\label{sec:tycoon}

In this section, we describe the \emph{Tycoon} design principles and
architecture and provide some preliminary simulation results
demonstrating its properties.

\subsection{Design Principles}

We use two design principles in the design of the \emph{Tycoon}
architecture: separation of mechanism and strategy and distribution of
allocation. Separation of mechanism and strategy is important because
they have different requirements and consequences for complexity,
security, and efficiency.

A strategy interprets a user's and an application's high level
preferences for how an application should be run into valuations of
resources. For example, web server may be more concerned with latency
than throughput and is therefore willing to consume a few resources on
many hosts in the hope one of its hosts will be close to a new
client. A database server or a rendering application is willing to
make a different tradeoff. Such preferences may not even be be
technical: an application distributing sensitive information may wish
to avoid hosts in certain countries. As a result of this diversity of
preferences, strategies that are specialized to particular users and
applications are more efficient than those that are not. However, if a
resource allocation system were to incorporate strategies as part of
its mechanism, it would either have to limit the preferences of
applications or increase the complexity of its mechanism. Examples of
the former approach are the system by Urgaonkar, et
al. \cite{urgaonkar2002}, which optimizes for throughout and
shortest-job-first allocation, which optimizes for latency.

A mechanism provides incentives for users to truthfully reveal their
values for resources and for providers to provide desirable
resources. The mechanism also needs to provide primitives for
expressing preferences. \emph{Tycoon} allows applications to specify
on which hosts they wish to run and the \emph{Auction Share} scheduler
allows them to specify how they wish to tradeoff throughput, latency,
and risk. The mechanism is critical to the security and efficiency of
the system, so it must be simple to understand and implement.

By separating strategy and mechanism, we allow the mechanism to be
simple while not limiting the preferences expressed by users and
applications. Instead, \emph{Tycoon} provides incentives for users and
application writers to specialize and optimize strategies. This
principle is similar to the original conception of how functionality
should be split between an operating system kernel and applications
(and was applied again with microkernels) and the end-to-end argument
\cite{saltzer1984endtoend} for how functionality should be split in
computer networks.

The other \emph{Tycoon} design principle is distribution of
allocation.  Since our motivation is to allocate resources for very
large systems like the Grid or PlanetLab, we distribute the allocation
of resources as much as possible (the bank is still centralized, as
described below).  This increases reliability because the failure of
one host will not prevent allocating resources on another. In
addition, distribution mitigates accidental or malicious misbehavior
by one host (e.g., charging credits without providing resources).
Users or parent agents (see below) will eventually notice that some
hosts have poor price/performance and run on other hosts. Finally,
distributed allocation reduces the latency to change allocations
because all allocation decisions are made local to a host.

\subsection{Architecture}

Using the principles described in the previous section, we split
\emph{Tycoon} into the following components: Parent Agent, Child
Agent, Auctioneer, Service Location Service, and Bank. The Parent
Agent and Child Agent implement the strategy, while the Auctioneer
implements the mechanism. The Service Location Service and the Bank
are infrastructure. 

\begin{CompactItemize}

\item \textbf{Parent Agent:} The parent agent does all high-level
distributed resource management on behalf of a user. Its two main
tasks are budgeting and managing child agents. Budgeting is
important because it removes the burden of managing budgets from the
user (at the cost of some flexibility). Parent agents should be
specialized for specific applications, but our current implementation
includes a sample parent agent for batch application. The user
specifies a number of credits, a deadline, and number of hosts to run
on. If the user specifies to spend \$700 for 100 minutes on seven
hosts, then the batch parent agent budgets \$1 for each host per
minute. 

Managing the child agents is important because some hosts may be more
cost-effective than others. This may be because heterogeneity in the
host platform or because one host is more lightly loaded than
another. The batch parent agent monitors progress and costs associated
with candidate hosts by querying the child agents. If a child agent
has a low performance to cost ratio, it kills the child agent and
associated application process running on that host. It replaces that
child agent with a randomly selected host in the hopes that it will
perform better. The key concern with this algorithm is the overhead
associated with copying code to a new host, especially if it is not
significantly better than the old candidates. We are still evaluating
the effectiveness of this algorithm.

\item \textbf{Child Agent:} Child agents bid for resources on hosts
and monitor application progress. Although we describe a child agent
as ``bidding'', a child agent actually transfers a lump sum to the
auctioneer which then does the fine-grained bidding itself (described
in more detail in \Section~\ref{sec:auctionshare}. This is more
efficient than communication between the child agent and the
auctioneer and removes the need to communicate frequently with the
bank. Child agents monitor application progress by maintaining
application specific statistics, e.g., the latency and throughput of
transactions on a web server or the rate of frames rendered for a
rendering application.

\item \textbf{Auctioneer:} Auctioneers schedule local resources in a
way that approximates proportional share, but allows flexibility for
latency-sensitive and risk-averse applications. Auctioneers do
efficient first or second price sealed bid auctions for fine-grained
resources, e.g., 10 ms CPU timeslices. This allows for high
utilization and the agility to adapt very quickly to changes in demand
and/or supply. We describe this mechanism in more detail in
\Section~\ref{sec:auctionshare}.

\item \textbf{Bank:} The bank maintains account balances for all users
and providers. The two key issues with the bank are security and
funding policy. The security problem is counterfeiting of currency. We
deal with this problem by only allowing transfers between
accounts. Users pay providers by directly transferring funds from one
account to another. This prevents counterfeiting, but involves the
bank in all transactions, which could limit scalability. We intend to
examine this, but we do not believe it will be a problem in practice
because (as described above) transfers only occur when a child agent
1) initially funds its application, 2) refreshes those funds when they
are exhausted, and 3) the parent agent's budget changes.

Funding policy determines how users obtain funds. We define \emph{open
loop} and \emph{closed loop} funding policies. In an open loop funding
policy, users receive an allotment of funds when they join and
at set intervals. The system administrators set their income rate
based on exogenously determined priorities. Providers accumulate funds
and return them to the system administrators. In a closed loop (or
\emph{peer-to-peer}) funding policy, users themselves bring resources
to the system when they join. They receive an initial allotment of
funds, but they do not receive funding grants after joining. Instead,
they must earn funds by enticing other users to pay for their
resources. A closed loop funding policy is preferable because it
encourages service providers to provide desirable resources and
therefore should result in higher economic efficiency.

\item \textbf{Service Location Service (SLS):} Parent agents use the
SLS to locate particular kinds of resources and auctioneers use it to
advertise resources. Although we currently use a simple centralized
soft-state server, we can use any of the distributed SLSs described in
the literature. The key point is that \emph{Tycoon} does not require
strong consistency. Parent agents monitor and optimize the end-to-end
performance of their applications, so stale information in the SLS
will simply delay the parent agent's from converging on an efficient
set of resources.

\end{CompactItemize}

\subsection{Results}

\begin{figure}[htb]
\includegraphics[angle=270,width=8.5cm,totalheight=5cm]{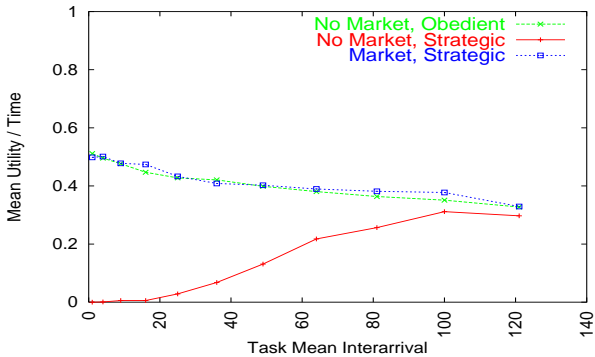}
\caption{\small The utility of different user behaviors and mechanism
  as a function of system load. }
\label{fig:market_no_market}
\end{figure}

We have preliminary simulation results. These results show that a
market for computational resources effectively maintains a high
utility despite strategic users. For the simulations in this section,
we use a market-based proportional share instead of the \emph{Auction
Share} described in \Section~\ref{sec:auctionshare}. We use a separate
simulator for the \emph{Auction Share} scheduler. This is our initial
approach because the \emph{Tycoon} simulator simulates events on many
hosts on large timescales while the \emph{Auction Share} simulator
simulates events on one host on small time scales. We are working on
merging the simulators. 

The simulation results are of 100 users submitting tasks with a
Poisson inter-arrival time. The simulation runs for 1000
seconds. There are ten hosts available for running tasks. We assume in
the simulation that there is no overhead for distributing and starting
tasks. The size and deadline of the tasks are also from a Poisson
distribution. Each task has a value selected uniformly at random from
$(0,1]$. If a task completes by the deadline, then the user receives
$value * size$ utility, otherwise nothing.  We simulate three
different user behaviors: obedient, strategic without a market, and
strategic with a market. Obedient users assign a Proportional Share
weight to their tasks equal to the task's value. Non-market strategic
users assign the maximum possible weight to all their tasks. Market
strategic users have a limited budget for assigning weights. These
users have an income of \$1 per time unit. They can save this income
or spend it by assigning some of it as the weight of one of their
tasks.

Market strategic users use a \emph{budgeting} strategy. They assign
weights at each host at each time unit to be
\[
\frac{balance * value}{num\_hosts * (deadline - now)}
\]
where $balance$ is the user's current credit balance, $value$ is the
value of the user's currently most valuable task, $num\_hosts$ is the
number of hosts to run on, $deadline$ is the deadline of the currently
most valuable task, and $now$ is the current time. 

Figure~\ref{fig:market_no_market} shows the simulation results. The
y-axis is the mean utility per host per time unit. This cannot exceed
1.0, but the only way for that to be the maximum is if there is always
a value 1.0 task in the system, which is not true in most cases. The
y-axis shows the mean inter-arrival of tasks in the system and is a
measure of overall system load. Each point in the graph is a run of
the simulator. As the load increases to the left, the obedient users
without a market are able to maintain a high level of utility.  In
contrast, the non-market strategic users are able to maintain a high
level of utility when the system is moderately loaded (from 120 to
100), but when the load saturates the system, utility drops to
zero. This is because the system wastes resources running tasks that
never meet their deadlines and therefore provide no utility. As the
number of tasks increases, this becomes more likely. In a system
without a mechanism or significant social pressure, this is
inevitable. With the market mechanism, the strategic users are forced
to truthfully reveal the value of their tasks and the system can
maintain a high utility.



\section{Auction Share Scheduling}
\label{sec:auctionshare}

In this section, we describe the \emph{Auction Share} scheduling
component of \emph{Tycoon} and use simulations to compare it with
Proportional Share scheduling. 

\subsection{Algorithm}

The \emph{Auction Share} scheduler achieves the high utilization of a
proportional share scheduler, the low latency of a Borrowed Virtual
Time scheduler, the low risk of reservations, and the
strategyproofness of a market scheduler. In addition, it is fair and
computationally efficient.

While we only describe the use of the auction scheduling algorithm for
CPU scheduling, it has a straightforward extension to other resources
like network bandwidth and disk storage. For CPU scheduling, the
resources are 10ms timeslices of the processor. The algorithm consists
of child agents that bid for resources for an application process and
an auctioneer that resolves the bids, allocates resources, and
collects credits. In a typical operating system like Linux, part of
the auctioneer resides in the kernel's processor scheduler.

Each child agent $i$ has a balance of $b_{i}$ credits, an expected
funding interval of $E(t_{i})$, and an expected number of
processor-seconds needed during $E(t_{i})$ of $q_{i}$. A parent agent
funds its child agents periodically in proportion to their importance
to it (\Section~\ref{sec:tycoon} describes this budgeting process in
more detail). $E(t_{i})$ is the average amount of time between such
fundings. We assume that $E(t_{i})$ is on the order of seconds and
therefore large relative to the timeslice size.

The child agent of a batch application sets $q_{i}$ to be $E(t_{i})$
in processor-seconds because batch application want to run as much as
possible.  The child agent of a delay-sensitive application sets
$q_{i}$ to be less than $E(t_{i})$ because the application is willing
to sacrifice some processor-seconds for lower delay. For example, a
web server is willing to sleep sometimes in return for having priority
when a request comes in. More willing an application is to trade
throughput for delay, the smaller its $q_{i}$ is relative to its
$E(t_{i})$.

To allocate a timeslice, the auctioneer computes the bid of each
thread $i$ as $b_{i}/q_{i}$. The auctioneer allocates the timeslice to
the thread with the highest bid.  After $elapsed$ elapsed seconds, the
running thread is context-switched either because its allocation
finished or because another thread with a higher bid becomes
runnable. At this point, the thread pays its bid to the auctioneer in
proportion to the amount of elapsed time:

\[
\frac{elapsed}{timeslice}*\frac{b_{i}}{q_{i}}
\]
The auctioneer then deducts this amount from the winning process's
balance. Alternatively, the auctioneer can charge the winning process
the second highest bidder's bid. We are still investigating the
tradeoffs of using the first or second price.

This algorithm is strategyproof because it corresponds to a series of
first or second price sealed bid auctions. The only difference is that
the \emph{Auction Share} auctioneer automatically computes bids for
the clients instead of having them do it. If they wish, clients can
micromanage the bidding by changing $q_{i}$, but only clients that
wish to change their latency-throughput tradeoff gain anything from
doing so.

\emph{Auction Share} is computationally efficient because the only
work the auctioneer needs to do each timeslice is update the previous
winning processes balance and select the highest (and possibly second
highest) current bid. The scheduler implementations of current
operating systems already do similar calculations at low overhead. A
typical implementation keeps process priorities in a heap which allows
the selection of the highest value in $O(1)$ time, and updating one of
the values in $O(log n)$ time, where $n$ is the number of
values. Changing $q_{i}$ and funding (which changes $b_{i}$) will also
require $O(log n)$, but these happen infrequently. 

This basic algorithm has high utilization, low latency,
strategyproofness, fairness, and low overhead, but it still has
significant risk. The arrival of new child agents will reduce the
resources allocated to all the other child agents using the
processor. Some risk-averse users would prefer having a higher
lower bound on the resources they receive in an interval instead of
having more total resources in that interval. An example is a
real-time process like a game server that would benefit more from
processing all its requests by their deadlines rather than finishing
some very quickly and some very slowly.

To satisfy these processes, \emph{Auction Share} offers a form of
reservation. The idea is to use recent history as a guide to calculate
a price for the reservation. A process can request a percentage of the
process $r$ for a time period of $p$ timeslices. In some cases, the
auctioneer must reject the reservation immediate because it has
already sold its limit of reservations. If this is not the case, the
auctioneer calculates the price for this reservation as
\[
(\mu + \sigma) * r * p
\]
where $\mu$ is the average price per timeslice, and $\sigma$ is the
standard deviation of the price. The process can either reject this
price, or pay it, in which case, $p$ begins immediately. During the
reservation, the auctioneer enters a proxy bid in its own auction such
that the reserving process always receives $r$ of the processor. 

This assumes the price in the recent past is indicative of the price
in the near future and that price is normally distributed. We are
still investigating methods for pricing reservations when these
assumptions do not hold. Another issue is that the auctioneer must
limit how much of the resource is reserved and the length of
reservations to maintain liquidity in its market. We have not
determined how these limits should be set.

\subsection{Results}
\label{sec:auctionshareresults}

\begin{table}[tbp]
\begin{center}
\begin{tabular}{|p{2.2cm}|c|p{.7cm}|p{1.2cm}|p{.95cm}|}
\hline
\scriptsize Algorithm &
\scriptsize Weight &
\scriptsize Yields CPU&
\scriptsize Scheduling Error&
\scriptsize Mean Latency\tabularnewline
\hline
\hline 
Proportional Share&
1/10&
yes&
0.09&
\small 81 ms\tabularnewline
\hline 
Proportional Share&
7/10&
yes&
0.01&
\small 4.4 ms\tabularnewline
\hline 
Proportional Share&
7/10&
no&
1.16&
\small 4.7 ms\tabularnewline
\hline 
Auction Share&
1/10&
yes&
0.01&
\small 3.6 ms\tabularnewline
\hline 
Auction Share&
1/10&
no&
0.02&
\small 96 ms\tabularnewline
\hline
\end{tabular}
\caption{\small The scheduling error and latency for different
  scheduling mechanisms with different application behaviors.}
\label{table:SimulationResults}
\end{center}
\end{table}

Our simulation results demonstrate that \emph{Auction Share} achieves
high utilization, low latency, and high fairness while providing an
incentive for truth-telling to rational users.  A proportional share
scheduler can achieve high utilization and either low latency or
fairness, but not both, and it does not provide incentives for
truth-telling. We simulate a latency sensitive application like a web
server running with 3 batch applications on a single processor. The
desired long term processor share for the web-serving application is
1/10. During each timeslice, the web server has a 10\% probability to
receive a request, which takes 10ms of CPU cycles to
service. Otherwise, the web server sleeps. The batch applications are
always ready to run. For the proportional share scheduler, we
initially set the weight of the web server and batch applications to
be 1, 2, 3, and 4, respectively. For the auction scheduler, we set the
income rates to be 1, 2, 3, and 4. For the auction scheduler, the
processes are not funded at precise intervals. Instead, the income
rates specify the mean interarrival times of funding. We run 1,000
timeslices of 10ms.

Table~\ref{table:SimulationResults} shows the latency and fairness for
different mechanisms and different application behaviors. ``Weight''
is the weight (for proportional share) or income rate (for auction
scheduling) given the web server. ``Yields CPU'' is whether the web
server correctly yields the CPU after servicing a
request. ``Scheduling Error'' measures by how much the actual CPU time
used by applications deviates from the amount intended. This is
computed as the sum of the relative errors for each of the
applications. For example, 0.09 indicates that the sum of the relative
errors is 9\%. Fairness is inversely proportional to the scheduling
error. ``Mean Latency'' is the mean latency for the latency-sensitive
application to service requests.

The second row of Table~\ref{table:SimulationResults} shows that
proportional share scheduling provides low error, but high
latency. Note that this latency is proportional to the total number of
runnable processes in the system, which is only four in our
simulations. We can reduce the latency by increasing the weight of the
web server, as shown in the third row. This assumes that the web
server yields the processor after finishing a request.  However, a
rational user will exploit the extra weight granted to his application
to do other computations to his benefit. Unfortunately, as shown in
the fourth row, this is at the expense of the overall fairness of the
system.

With auction share scheduling, the weight of the web server does not
need to be boosted to achieve low latency (as shown in the fifth
row). More importantly, if the web server mis-estimates the resources
it requires (accidentally or deliberately), as shown in the last row,
it only penalizes its own latency. The overall fairness of the system
remains high. This provides the incentive for child agents to
truthfully reveal their requirements for resources and therefore
allows the system to achieve high economic efficiency. In addition,
\emph{Auction Share} has the same utilization as Proportional Share
because the processor is always utilized.

\section{Conclusion}
\label{sec:conclusion}

The contributions of this paper are the \emph{Tycoon} distributed
market-based resource allocation architecture and the \emph{Auction
  Share} local resource scheduler. Through simulation, we show that a
market-based system has greater utility than a non-market-based
proportional share system. In addition, we show through simulation
that \emph{Auction Share} is high utilization, low latency, and fair. 

We are planning to implement \emph{Auction Share} in the Linux
kernel. Using this, we hope to deploy our implementation of
\emph{Tycoon} on a large cluster and take measurements of realistic
workloads.




{\footnotesize
\bibliographystyle{acm}
\bibliography{bibliographies/resource_allocation,bibliographies/economics,bibliographies/networking,bibliographies/overlay,bibliographies/network_performance,bibliographies/peer-to-peer,bibliographies/security,bibliographies/reputation,bibliographies/network_architecture}
}

\end{sloppypar}

\end{document}